\documentclass[12pt]{iopart}

\usepackage{multirow}
\usepackage{graphicx}
\begin{document}

\title[A generalized theory of preferential linking]{A generalized theory of preferential linking}

\author{Hai-Bo Hu$^1$, Jin-Li Guo$^2$ and Xuan Liu$^1$}

\address{$^1$East China University of Science and Technology, Shanghai 200237, China\\
$^2$University of Shanghai for Science and Technology, Shanghai 200093, China}
\ead{sdhuzi@163.com}
\begin{abstract}
There are diverse mechanisms driving the evolution of social
networks. A key open question dealing with understanding their
evolution is: How various preferential linking mechanisms produce
networks with different features? In this paper we first empirically
study preferential linking phenomena in an evolving online social
network, find and validate the linear preference. We propose an
analyzable model which captures the real growth process of the
network and reveals the underlying mechanism dominating its
evolution. Furthermore based on preferential linking we propose a
generalized model reproducing the evolution of online social
networks, present unified analytical results describing network
characteristics for 27 preference scenarios, and explore the
relation between preferential linking mechanism and network
features. We find that within the framework of preferential linking
analytical degree distributions can only be the combinations of
finite kinds of functions which are related to rational, logarithmic
and inverse tangent functions, and extremely complex network
structure will emerge even for very simple sublinear preferential
linking. This work not only provides a verifiable origin for the
emergence of various network characteristics in social networks, but bridges the micro individuals' behaviors and the global organization of social networks.

\end{abstract}

\maketitle

\section{Introduction}

In real life not everyone is equally popular, and in social networks
also not everyone possesses the same status or position. Some
individuals tend to be at the center of social networks while others
remain on the periphery [1, 2]. This realization gave rise to the concept
of network centrality [3]. Centrality has important
effects on the evolution of social networks. Degree centrality, i.e.
the number of ties that an actor possesses, has received particular
attention maybe due to its computational simplicity. In many
real-world social networks, researchers have found that most actors
have only a few ties, while a small number have extraordinarily
many. For instance it was found that degree
distribution is highly skewed in sexual contact networks, where some
super-connecter actors acquire as many as 1000 partners [4]. Similar
patterns also exist in movie co-appearance network, and numerous
co-authorship networks in academia [5].

In the past few years, Web 2.0 which is characterized by social
collaborative technologies, such as social networking site (SNS),
blog, Wiki, video or photo sharing and folksonomy, has attracted
much attention of researchers from diverse disciplines [6]. As a fast growing business, many SNSs of different
scopes and purposes have emerged on the Web [7], many of
which, such as \emph{Facebook} [8], \emph{Renren} [9], \emph{MySpace} [10, 11],
\emph{Orkut} [10, 12] and newborn \emph{Google+} [13], are among the most popular sites on the Web. Users
of these sites, by establishing friendship relations with other
users, can form online social networks (OSNs). Like real-world
social networks, in OSNs individual degrees also show obvious
heterogeneity. An analysis of the 721 million users on
\emph{Facebook} found that a few individuals have 5000 friends (a
limit imposed by \emph{Facebook}), more than 26 times as many as the
average user's 190 [14].

One important reason social networks develop such a high variance in
actors' degrees is that the number of ties an actor possesses
affects processes of attachment. Social connections tend to accrue
to those who already have them, the consequence of which is that
small differences in actor degree compound over time into a distinct
cumulative advantage [15, 16]. In OSNs the creation of
links between individual users has been studied in a number of
contexts [17, 18], and is believed to be driven
by the principle of preferential attachment (PA), i.e. new users
prefer to connect to old users with higher degree. PA is widely
recognized as the principal driving force behind the evolution of
many growing networks. Besides the PA hypothesis stands as the
accepted explanation behind the prevalence of scale-free
organization in diverse evolving networks.

That to what extent PA works has been studied, qualitatively or
quantitatively, in real-world and OSNs. However
most of the researches are empirical and lack analyzable models.
Besides in network evolution when new users establish friend
relationship with old users, or new ties are established between old
users, the old users with large degrees are all likely to be
preferentially selected. However most previous researches either
only focus on PA or combine the two cases into one, overlooking
possible preference of varying degrees for link establishment under
different scenarios. To date, there are few analytical studies that
bridge the micro preferential linking (PL, considering link
establishment not only between old users and new users but between
old users) and macrostructure of OSNs. A key open question dealing
with understanding the evolution of OSNs is: How will the
combination of linear PL, sublinear PL and randomized attachment
generate networks with different characteristics? In
this paper we exploit not only how linear PL leads to networks with scale-free feature (which has been partly
studied in the past), but also what network features will result
from diverse PL mechanisms, which has not been previously studied.

In the reminder of this paper, after an overview of PA in social networks, we present a detailed case study based on real network dataset, following the procedure of network measurement,
modeling, analysis, and model validation. We bring forward an
analyzable model, which can reproduce the process of network growth
and connect the PL mechanism and the network characteristics. Furthermore
considering different forms of PL, we propose a
generalized model for the evolution of OSNs, and present analytical results characterizing network features for
diverse preference scenarios. At last from the perspective of sociology and
economics we analyze the reasons why PL exists in OSNs. We discuss the limitation
of the paper and a research framework for better understanding
the evolution of OSNs is presented.

\section{Preferential Attachment}

Many social networks have a measured degree distribution $P(k)$ that
is either a power-law $P(k) \propto {k^{ - \gamma }}$, or a
power-law with an exponential cutoff. Growing models have been proposed to
account for these features, most of them being based on
some form of PA. Generally PA means that when new nodes join the network linking to the existing nodes,
the probability of linking $i$ is an increasing function of the degree
$k_i$ of $i$. Some models assume this function to be linear [19], while in other cases it has been assumed to
depend on a different power of $k_i$ [20]. In
general, we have that the probability $\Pi ({k_i})$ with which an
edge belonging to a new node connects to an existing node $i$ of
degree $k_i$ will be $\Pi {({k_i})} \propto k_i^\beta $, where $\beta \ge 0$.
For $\beta=1$ the rate is linear and the model reduces to the
familiar BA model which yields a power-law
degree distribution with $\gamma = 3$ [19]. For
$\beta<1$ the PA is sublinear and $P(k)$ is a
stretched exponential $P(k) \propto k^{ - \gamma } \exp \left[ { - (b(\gamma )/(1 - \gamma ))k^{1 - \gamma } } \right]$, where $b$ is a
constant depending on $\gamma$ [20]. The absence of
PA is attained in the limit $\beta= 0$, when
the attachment rule is independent of degree. The resulting degree
distribution in this case is given by $P(k) \propto \exp ( - k/m)$ where $m$ is a
constant. For $\beta>1$ a single node gets almost all the edges, with the rest
having an exponential distribution of the degrees. Therefore, to
know which kind of PA, if any, is at work in a
particular growing network, one needs to study empirically networks
for which the time at which new nodes entered the network and new
edges formed is known.

In recent years some empirical researches have verified the
existence of a PA rule for social networks, including real-world and
online, and exponent $\beta$ has also been estimated for several
networks. However there are some differences as for the functional
form of $\Pi {({k_i})}$. In some cases it appears to be quite close
to linear, while in other cases it has been found to be sublinear.

For real-world social networks, Newman studied
scientific collaboration networks and found that researchers in
physics and biology who already had a large number of collaborators
are more likely to accumulate new collaborators in the future [21]. By
fitting data he obtained $\beta = 1.04$ for Medline and $\beta =
0.89$ for the Los Alamos Archive. Jeong et al. explored the
co-authorship network in the neuroscience field and the Hollywood
co-cast actor network, and found that $\beta=0.79$ for the
co-authorship network and $\beta=0.81$ for the co-cast actor
network, implying sublinear PA [22]. Peltom\"{a}ki and Alava studied
growing collaboration networks from the IMDB and arXiv.org preprint
server, and found that for the actor network the measured value of
the exponent $\beta \approx 0.65$, for the astrophysics network
$\beta \approx 0.6$, and for the condensed matter physics and high
energy physics networks $\beta \approx 0.75$ [23]. de Blasio et al. tested the PA conjecture in sexual contact networks based on Norwegian survey data , and found evidence of nonrandom, sublinear
PA [24].

Recently due to the availability of data of evolving OSNs though
they may be low-resolution or only a sample during a period of time,
PA mechanism has also been validated in OSNs. Mislove et al. studied the evolution of \emph{Flickr} and found that users tend to create and receive links in proportion to their outdegree and
indegree, respectively [25]. Leskovec et al. studied the evolution
of \emph{Flickr}, \emph{del.icio.us}, \emph{Yahoo!Answers} and
\emph{LinkedIn}, and examined whether PA holds for the networks [26].
They found that \emph{Flickr} and \emph{del.icio.us} show linear
preference, and \emph{Yahoo!Answers} shows slightly sublinear
preference, $\beta=0.9$. For \emph{LinkedIn} for low degrees,
$\beta=0.6$; however, for large degrees, $\beta=1.2$, indicating
superlinear preference. Garg et al. analyzed an evolving online
social aggregator \emph{FriendFeed} and found that for source node
selection $\beta=0.8$ and for destination node selection,
$\beta=0.9$ [27]. Szell and Thurner studied a massive multiplayer online
game \emph{Pardus} [28]. They measured indegrees of characters who are
marked by newcomers as friend (enemy) and found that $\beta = 0.62$
for friend markings with $k_{\rm{in}} < 30$, and $\beta= 0.90$ for
all enemy markings. Aiello et al. investigated the dynamical
properties of \emph{aNobii} and tested PA mechanism [29]. They obtained a
linear behavior, both when considering for $k$ the in and the
outdegree. Rocha et al. studied the sexual networks of
Internet-mediated prostitution extracted from a forum-like Brazilian
Web community and found that sex-buyers exhibit sublinear PA for
both short and long intervals [30]. They also observed close to linear PA
for sex sellers for short time intervals, whereas longer time
intervals are associated with sublinear PA. This means that feedback
processes are stronger for shorter than for longer timescales.
Moreover Zhao et al. studied the evolution of \emph{Renren},
the largest OSN in China, and found that $\beta$ is not a constant
over time [9]. $\beta (t)$ decreases as the network grows which
indicates that the influence of PA on network evolution weakens with
the growth of \emph{Renren}.

From the previous theoretical and empirical researches we find that
although the basic idea of PA is already well established, the relation
between the combination of various PL mechanisms and resulting network
features has not been fully exploited, which is the primary goal of the
paper.

\section{Case Study}
\subsection{Dataset}

Uncovering how the micro-mechanisms of network growth lead to the
macrostructure of OSNs is of paramount importance in understanding
the evolution of OSNs; however data privacy policy makes it
difficult for researchers to obtain the data of evolving OSNs. Thus it is very difficult to capture the process of
network evolution due to the fact that detailed empirical data of
network growth with time labels integrating the joining of new users
and establishment of new friend relationship are still scarce.
Although some works studied growing OSNs like \emph{Facebook} [31] and \emph{Renren} [9], the datasets studied do not indicate who is sender and who is receiver for a link request.

In this section we first study \emph{Wealink}, a large
\emph{LinkedIn}-like SNS whose users are mostly
professionals, typically businessmen and office clerks. The network
data, logged from 0:00:00 h on 11 May 2005 (the inception day for
the Web 2.0 site) to 15:23:42 h on 22 August 2007, include all
friend relationship and the time of formation of each tie.

The finial data format, as shown in Fig. 1, is a time-ordered list
of triples $<$$U_i$, $U_j$, $T_k$$>$ indicating that at time $T_k$
user $U_i$ sends a link request to user $U_j$ or $U_i$ accepts
$U_j$'s previous friendship request and they become friends. Like
\emph{Facebook} and \emph{Renren} only when the sent invitations are
accepted will the friend relations be established. The online
community is a dynamically evolving one with new users joining the
network and new ties established between users.

\begin{figure}
\centerline{\includegraphics[height=1.5in]{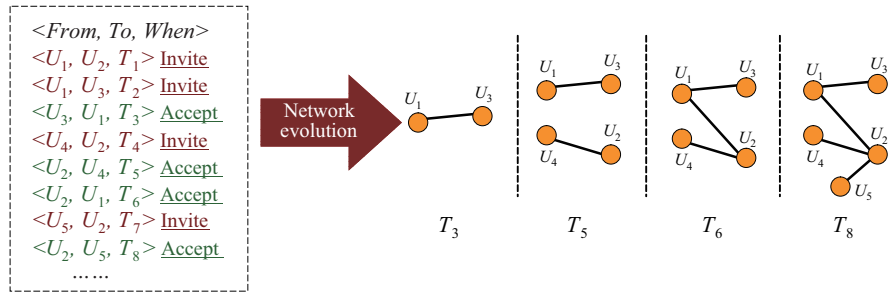}} \caption{Data
format and evolution of OSN \emph{Wealink}.} \label{fig:one}
\end{figure}

\subsection{Preferential Linking}

Like some other OSNs the degree distribution of \emph{Wealink} shows
power-law feature. This kind of distribution can be produced through
linear PA, as revealed by BA model. In addition to the dynamics that
is due to new users joining the network (generally by creating a new
account) and making friends with the old users, there is also the
dynamics that results from active users interacting with each other.
In real scenario of network growth when new users establish friend
relationship with old users, or new ties are established between old
users, the old users with large degrees are all likely to be
preferentially selected. In this subsection we will give evidence supporting these hypotheses.

Since many OSNs are consequence of bilateral decisions of a pair of
users, not of their unilateral decisions, to test the preference
feature for different types of link establishment, we separate PL into three aspects: preferential acceptance, preferential
creation, and PA. Preferential acceptance implies that, the larger
an old user's degree is, the more likely she/he will be selected as
friends by the other old users. Preferential creation implies that,
the larger an old user's degree is, the more likely her/his link
invitations will be accepted by the other old users. The meaning of
PA remains unchanged, i.e. new users tend to attach to already
popular old users with large degrees.

Let $k_i$ be the degree of user $i$. The probability that user $i$
with degree $k_i$ is chosen can be expressed as
\begin{equation}
  \prod (k_i ) = \frac{{k_i^\beta  }}{{\sum\nolimits_j {k_j^\beta  }
  }}.
  \label{1}
\end{equation}
We can compute the probability $\Pi (k)$ that an old user of degree
$k$ is chosen, and it is normalized by the number of users of degree
$k$ that exist just before this step:
\begin{equation}
  \prod (k) = \frac{{\sum\nolimits_t {\left[ {e_t  = v \wedge k_v (t - 1) = k} \right]} }}{{\sum\nolimits_t {\left| {\left\{ {u:k_u (t - 1) = k} \right\}} \right|} }} \propto
  k^\beta,
  \label{2}
\end{equation}
where $e_t  = v \wedge k_v (t - 1) = k$ represents that at time $t$
the old user whose degree is $k$ at time $t-1$ is chosen. We use $[
\cdot ]$ to denote a predicate (which takes a value of 1 if the
expression is true, else 0). Generally, $\Pi (k)$ has significant
fluctuations, particularly for large $k$. To reduce the noise level,
instead of $\Pi (k)$, we study the cumulative function:
\begin{equation}
  \kappa (k) = \int_0^k {\prod (k){\rm{d}}k}  \propto k^{\beta  + 1} = k^{\alpha}.
  \label{3}
\end{equation}

Fig. 2 shows the relation between degree $k$ of users and preference
metric $\kappa$. Least squares linear regression gives $\alpha =
1.93\pm0.01 (R^2=0.99)$ for preferential creation, $\alpha =
1.97\pm0.01 (R^2=0.98)$ for PA and $\alpha = 2.06\pm0.01 (R^2=0.99)$
for preferential acceptance. All are with significance level
$p<2.2 \times 10^{ - 16}$. Thus $\beta \approx 1$ indicating linear
preference.

\begin{figure}
\centerline{\includegraphics[width=3in]{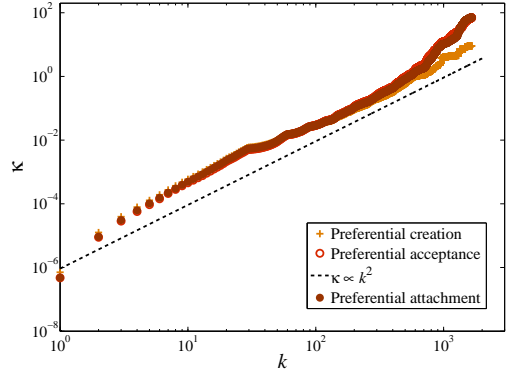}}
\caption{Preference characteristics in the evolution of
\emph{Wealink}.} \label{fig:two}
\end{figure}

\subsection{Model}

Like other OSNs the evolution of \emph{Wealink} includes two
processes. The first one is that a new user joins in the network and
establishes friend relation with an old user already present in the
network. The second one is that a friend relation is established
between two old users. Certainly there exists the case that a tie
forms between two new users; however the situation is rare in real
world and can be neglected.

Based on the linear preference we bring forward the following
network model. Starting with a small connected network with $m_0$
users, at every time step, there are two alternatives:

$A$. With probability $p$, we add a new user with one edge that will
be connected to the user already present in the network. The
probability that the new user will be connected to old user $i$ with
degree $k_i$ is $\Pi ({k_i}) = {k_i}/\sum\nolimits_j {{k_j}}$.

$B$. With probability $q=1-p$, we add one new edge connecting the
old users. The two endpoints of the edge are also chosen according
to linear preference.

After $t$ time steps the model leads to a network with mean number
of users $N(t)  = m_0  + pt$. For large $t$, $N \approx pt$ and the
total degree of the network $k_{{\rm{all}}}(t) \approx 2t$. Applying
mean-field approach for user $i$, we obtain
\begin{equation}
\frac{{\partial {k_i}}}{{\partial t}} =
p\frac{{{k_i}}}{{\sum\nolimits_j {{k_j}} }} +
2q\frac{{{k_i}}}{{\sum\nolimits_j {{k_j}} }} = \frac{{p +
2q}}{{2t}}{k_i}. \label{1}
\end{equation}
The solution of Eq. (4) with the initial condition $k_i(t_i)=1$ is
\begin{equation}
{k_i} = {(t/{t_i})^{\frac{{p + 2q}}{2}}}. \label{2}
\end{equation}
Thus
\begin{equation}
P({k_i} < k) = P({t_i} > {k^{-\frac{2}{{p + 2q}}}} \cdot t).
\label{3}
\end{equation}
The probability density of $t_i$ for large $t$ is
\begin{equation}
{P_i}({t_i}) = {1 \mathord{\left/
 {\vphantom {1 {\left( {{m_0} + tp} \right)}}} \right.
 \kern-\nulldelimiterspace} {\left( {{m_0} + tp} \right)}} \approx 1/(tp). \label{4}
\end{equation}
From Eq. (6) we obtain
\begin{equation}
P({k_i} < k) = 1 - P({t_i} \le {k^{-\frac{2}{{p + 2q}}}} \cdot t)
= 1 - {p^{ - 1}} \cdot {k^{ - \frac{2}{{p +
2q}}}}. \label{5}
\end{equation}
Thus the probability density for $P(k)$ is
\begin{equation}
P(k) = \frac{{\partial P({k_i} < k)}}{{\partial k}} \propto k^{ - \frac{{4 - p}}{{2 - p}}}.
\label{6}
\end{equation}
The exponent $\gamma  \in (2,3]$ and when $p=1$ the model is reduced
to BA model.

According to empirical data, we obtain $p =  {\rm{0}}{\rm{.7941}}$
and $q=0.1939$. The links created between two new users are few and
thus can be negligible. Based on the parameters and Eq. (9), we obtain
$P(k)\propto k^ {-2.67}$. Fig. 3 shows the numerical result which is
obtained by averaging over 10 independent realizations with $p
={\rm{0}}{\rm{.7941}}$ and the same number of users as
\emph{Wealink}. Its degree exponent 2.62 agrees well with the
predicted value of 2.67. Fig. 3 also presents the complementary
cumulative degree distribution of \emph{Wealink}. We fit the network
data with power-law model utilizing Maximum Likelihood Estimate
method and obtain $\gamma = 2.91$. The predicted value of
the degree exponent 2.67 of the model achieves proper agreement with
the real value 2.91. We also compute $p$-value for the estimated
power-law fit to the network implementing the Kolmogorov-Smirnov
test and obtain $p=0.704$ [32]. We choose threshold
0.1, and thus the power-law fit is a good match to the degree
distribution of $Wealink$.

\begin{figure}
\centerline{\includegraphics[width=3in]{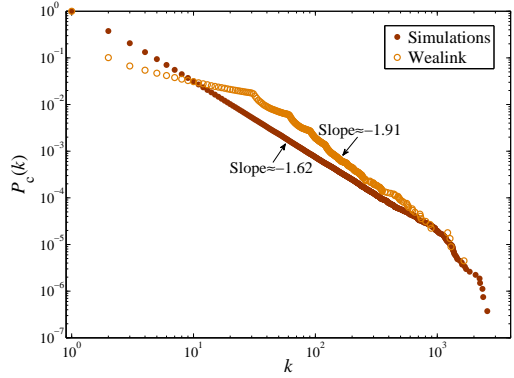}} \caption{The
complementary cumulative degree distributions of $Wealink$ and the
networks obtained by numerical simulations.} \label{fig:three}
\end{figure}

In real world different from the ideal model, the probability $p$
cannot be stationary during the evolution of OSNs. In some stage $p$
can be very large while in another stage $p$ can be very small,
which can lead to the difference between real exponent and predicted
one. Fig. 4 shows the evolution of $p$ and $q$, and demonstrates the
fact. As a guide we also indicate the positions of $p = 0.7941$ and
$q = 0.1939$.

\begin{figure}
\centerline{\includegraphics[width=3in]{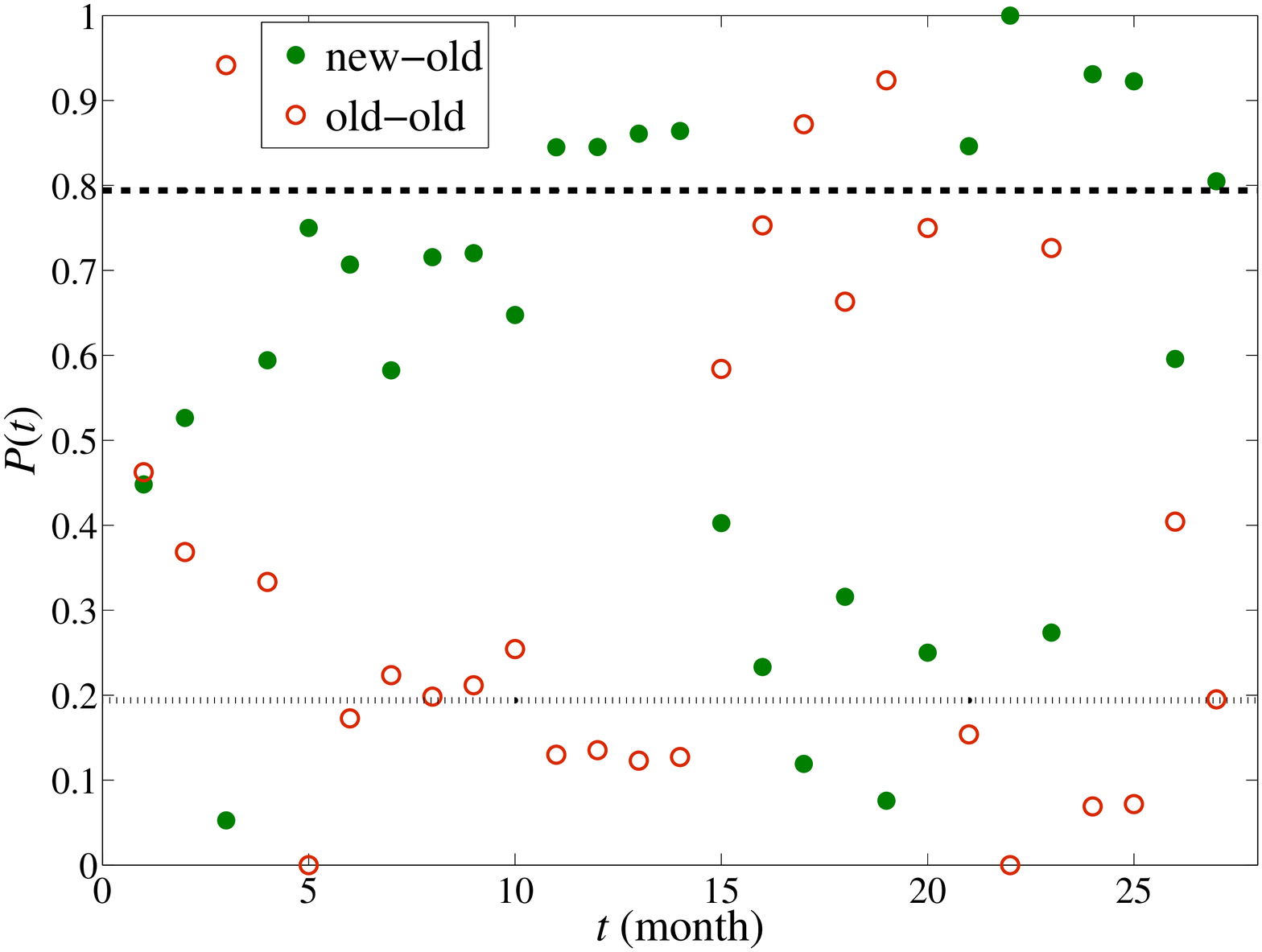}}
\caption{Evolution of the fraction of two kinds of edges. Dashed
line indicates $p = 0.7941$ while dotted line $q = 0.1939$.}
\label{fig:four}
\end{figure}

%Apparently, a larger $\beta$ leads to a greater heterogeneity
%of the network structure. For demonstration Fig. 5 shows the numerical
%result which is obtained by averaging over 10 independent
%realizations. If rule $B$ of the model is unchanged and in rule $A$
%we suppose that the probability that the new user will be
%connected to old user $i$ with degree $k_i$ is $\Pi ({k_i}) \propto
%k_i^\beta$, the corresponding numerical result for different $\beta$
%is shown in Fig. 5(a). If rule $A$ is unchanged and in rule $B$ we
%suppose that one endpoint is chosen according to $\Pi ({k_i}) \propto
%k_i^\beta$, the corresponding numerical result is shown in Fig.
%5(b). We find that when $\beta$ increases, ${P_{\rm{c}}}(k)$ gets
%more and more decentralized and hubs emerge for larger $\beta$.
%
%
%\begin{figure}
%\centerline{\includegraphics[width=5in]{fig5.eps}} \caption{The
%complementary cumulative degree distributions of networks obtained
%by numerical simulations with $p = 0.6$ and $N=10^5$.}
%\label{fig:five}
%\end{figure}

\section{Generalized Model}

In ONSs new users are constantly joining the social networks, and
create edges towards already present users. Very few users leave the
network, and very few edges disappear between users which remain in
the network. Edges on the other hand are created between already
present users. Besides in the evolution of real OSNs, new users or
edges are added into networks one by one, and previous empirical
researches have also shown that in OSNs most preference exponent
$\beta \le 1$. Thus we bring forward the following general network
model. Still starting with a small connected
network with $m_0$ users, however at every time step, there are
another two alternatives:

%\begin{figure}
%\centerline{\includegraphics[width=2.5in]{fig6.eps}} \caption{A
%general network model.} \label{fig:six}
%\end{figure}

$A$. With probability $p$, we add a new user with one edge that will
be connected to the user already present in the network. The
probability that the new user will be connected to old user $a$ with
degree $k_a$ is $\Pi ({k_a}) = k_a^\alpha /\sum\nolimits_j
{k_j^\alpha } $, where $0\le \alpha \le 1$.

$B$. With probability $q=1-p$, we add one new edge connecting the
old users. One endpoint $b$ is chosen according to $\Pi ({k_b}) =
k_b^\beta /\sum\nolimits_j {k_j^\beta } $ while another endpoint $c$ is
chosen according to $\Pi ({k_c}) = k_c^\gamma /\sum\nolimits_j
{k_j^\gamma } $, , where $0\le \beta, \gamma \le 1$.

Thus
\begin{equation}
\frac{{\partial {k_i}}}{{\partial t}} = p\frac{{k_i^\alpha
}}{{\sum\nolimits_j {k_j^\alpha } }} + q\frac{{k_i^\beta
}}{{\sum\nolimits_j {k_j^\beta } }} + q\frac{{k_i^\gamma
}}{{\sum\nolimits_j {k_j^\gamma } }}, \label{6}
\end{equation}
where $0 < p,q < 1$.

According to
\begin{equation}
\left\{ \begin{array}{l}
 \sum\nolimits_j {k_j^0  = pt}  \\
 \sum\nolimits_j {k_j^1  = 2t}  \\
 \end{array} \right.,
\end{equation}
when $0<\alpha<1$, $\sum\nolimits_j {k_j^\alpha   = ut} $ where
$p<u<2$.

As users $a$, $b$ and $c$ can be chosen according to any one of
three rules--random attachment, linear PL and sublinear PL, there
are 27 different scenarios for the evolution of ONSs.

First we consider the situations where only linear PL or random
attachment exists, i.e. $\alpha ,\beta ,\gamma=1$ or 0, and there
are totally eight scenarios which can be divided into six cases.
Utilizing the similar approach in Sec. 3, we get all their degree
distributions which have been summarized in Tab. 1. It is not
surprising that for case I linear PL will result in power-law
distribution, and for case VI random attachment will lead to
exponential distribution. However it is interesting that for the
other cases, the combination of linear PL component and randomized
attachment component also will generate networks with
approximatively power-law distribution. Besides according to the variation range of degree exponent in Tab. 1, obviously the introduction of randomized attachment can enhance the homogeneity of network structure.

\begin{table}[t]
\caption{\label{tab1}The evolution of $k_i$ and corresponding $P(k)$ when only linear PL or random attachment exists.}
\begin{indented}
\item[]\begin{tabular}{@{}llllll}
\br
Case & $a$ & $b$ & $c$ & $\partial {k_i}/\partial t$ & $P(k)$\\
\mr
I & Linear & Linear & Linear & $\frac{{p +
2q}}{{2t}}{k_i}$ & $\propto k^{ - \frac{{4 - p}}{{2 - p}}}$\\
\multirow{2}{*}{II}& Linear & Linear & Random  &
\multirow{2}{*}{$\frac{{{k_i}}}{{2t}} + \frac{q}{{pt}}$}&
\multirow{2}{*}{$\propto(kp+2q)^{-3}$}\\  & Linear & Random & Linear & & \\
III & Linear & Random & Random & $\frac{{p{k_i}}}{{2t}} + \frac{{2q}}{{pt}}$
& $\propto {(k{p^2} + 4q)^{ - \left( {1 + \frac{2}{p}} \right)}}$ \\
IV & Random & Linear & Linear & $\frac{1}{t} + \frac{{q{k_i}}}{t}$ & $\propto {(kq
+ 1)^{ - \left( {1 + \frac{1}{q}} \right)}}$ \\
\multirow{2}{*}{V}& Random & Linear & Random  & \multirow{2}{*}{$\frac{1}{{pt}} +
\frac{{q{k_i}}}{{2t}}$}&
\multirow{2}{*}{$\propto {(kpq + 2)^{ - \left( {1 + \frac{2}{q}} \right)}}$}\\  & Random & Random & Linear & & \\
VI & Random & Random & Random & $\frac{{p + 2q}}{{pt}}$ & $ \propto
{{\rm{e}}^{ - \frac{{pk}}{{p + 2q}}}}$ \\
\br
\end{tabular}
\end{indented}
\end{table}

When sublinear PL exists, there are 19 different scenarios for the
evolution of $k_i$ which can be divided into 12 cases and are shown
in Tab. 2. According to Lipschitz conditions there are unique
solutions to $k_i$.

\begin{table}
\caption{\label{tab2}The evolution of $k_i$ when sublinear PL exists. $0<\alpha, \beta, \gamma<1$ and $p<u, v, w<2$.}
\begin{indented}
\item[]\begin{tabular}{@{}lllll}
\br
Case & $a$ & $b$ & $c$ & $\partial {k_i}/\partial t$ \\
\mr
I & Sublinear & Linear & Linear & $\frac{{pk_i^\alpha }}{{ut}} + \frac{{q{k_i}}}{t}$ \\
\multirow{2}{*}{II}& Linear & Sublinear & Linear  & \multirow{2}{*}{$\frac{{qk_i^\beta }}{{vt}} + \frac{{{k_i}}}{{2t}}$}\\  & Linear & Linear & Sublinear &  \\
III & Sublinear & Sublinear & Sublinear & $\frac{{pk_i^\alpha }}{{ut}} + \frac{{qk_i^\beta }}{{vt}} + \frac{{qk_i^\gamma }}{{wt}}$ \\
IV & Random & Sublinear & Sublinear & $\frac{1}{t} + \frac{{qk_i^\beta }}{{vt}} + \frac{{qk_i^\gamma }}{{wt}}$ \\
\multirow{2}{*}{V}& Sublinear & Sublinear & Random  & \multirow{2}{*}{$\frac{{pk_i^\alpha }}{{ut}} + \frac{{qk_i^\beta }}{{vt}} + \frac{q}{{pt}}$}\\  & Sublinear & Random & Sublinear &  \\
VI & Linear & Sublinear & Sublinear & $\frac{{p{k_i}}}{{2t}} + \frac{{qk_i^\beta }}{{vt}} + \frac{{qk_i^\gamma }}{{wt}}$ \\
\multirow{2}{*}{VII}& Sublinear & Sublinear & Linear  & \multirow{2}{*}{$\frac{{pk_i^\alpha }}{{ut}} + \frac{{qk_i^\beta }}{{vt}} + \frac{{q{k_i}}}{{2t}}$}\\  & Sublinear & Linear & Sublinear &  \\
\multirow{2}{*}{VIII}& Random & Sublinear & Random  & \multirow{2}{*}{$\frac{1}{t} + \frac{{qk_i^\beta }}{{vt}} + \frac{q}{{pt}}$}\\  & Random & Random & Sublinear &  \\
IX & Sublinear & Random & Random & $\frac{{pk_i^\alpha }}{{ut}} + \frac{{2q}}{{pt}}$ \\
\multirow{2}{*}{X}& Linear & Sublinear & Random  & \multirow{2}{*}{$\frac{{p{k_i}}}{{2t}} + \frac{{qk_i^\beta }}{{vt}} + \frac{q}{{pt}}$}\\  & Linear & Random & Sublinear &  \\
\multirow{2}{*}{XI}& Random & Sublinear & Linear  & \multirow{2}{*}{$\frac{1}{t} + \frac{{qk_i^\beta }}{{vt}} + \frac{{q{k_i}}}{{2t}}$}\\  & Random & Linear & Sublinear &  \\
\multirow{2}{*}{XII}& Sublinear & Linear & Random  & \multirow{2}{*}{$\frac{{pk_i^\alpha }}{{ut}} + \frac{q}{{pt}} + \frac{{q{k_i}}}{{2t}}$}\\  & Sublinear & Random & Linear &  \\
\br
\end{tabular}
\end{indented}
\end{table}

For case I we obtain
\begin{equation}
\frac{{\partial k_i }}{{\partial t}} - \frac{q}{t}k_i  = \frac{p}{{ut}}k_i^\alpha,
\end{equation}
which is Bernoulli's differential equation.
Let $z = k_i^{1 - \alpha } $, thus
\begin{equation}
\frac{{\partial z}}{{\partial t}} - \frac{q}{t}(1 - \alpha )z = (1 - \alpha )\frac{p}{{ut}}.
\end{equation}
Therefore
\begin{eqnarray}
z & = & {\rm{e}}^{{\rm{ - }}\int {\frac{{(\alpha  - 1)q}}{t}{\rm{d}}t} } \left( {c + \int {\frac{{p(1 - \alpha )}}{{ut}}{\rm{e}}^{\int {\frac{{(\alpha  - 1)q}}{t}{\rm{d}}t} } {\rm{d}}t} } \right)\nonumber
\\
& = &  c_1 t^{(1 - \alpha )q}  - \frac{p}{{uq}},
\end{eqnarray}
where $c$ and $c_1$ are constants.
Thus
\begin{equation}
k_i  = \left[ {c_1 t^{(1 - \alpha )q}  - \frac{p}{{uq}}} \right]^{\frac{1}{{1 - \alpha }}} .
\end{equation}
According to initial value $k_i(t_i)=1$, we obtain
\begin{equation}
k_i  = \left[ {\left( {1  + \frac{p}{{uq}}} \right)\left( {\frac{t}{{t_i }}} \right)^{(1 - \alpha )q}  - \frac{p}{{uq}}} \right]^{\frac{1}{{1 - \alpha }}}.
\end{equation}
Accordingly
\begin{equation}
P(k) \propto \left( {uqk^{1 - \alpha }  + p} \right)^{ - \left[ {1 + \frac{1}{{(1 - \alpha )q}}} \right]},
\end{equation}
and for large $k$, $P(k) \propto k^{ - \left( {1 - \alpha  + \frac{1}{q}} \right)}$.
%The numerical simulation result of this case has been shown in Fig. 5(a).

Similarly for case II we obtain
\begin{equation}
P(k) \propto \left( {\frac{1}{2}vk^{1 - \alpha }  + q} \right)^{ - \left( {\frac{2}{{1 - \alpha }} + 1} \right)},
\end{equation}
and for large $k$, $P(k) \propto k^{ - (3 - \alpha )} $.
%The numerical simulation result of this case has also been shown in Fig. 5(b).

For case III when $\alpha=\beta=\gamma$
\begin{equation}
\frac{{\partial k_i }}{{\partial t}} = \frac{{(1 + q)k_i^\alpha  }}{{ut}},
\end{equation}
thus
\begin{equation}
k_i  = \left[ {1  + \frac{{(1 - \alpha )(1 + q)\ln \frac{t}{{t_i }}}}{u}} \right]^{\frac{1}{{1 - \alpha }}}.
\end{equation}
Accordingly
\begin{equation}
P(k) \propto k^{ - \alpha } \exp \left[ {\frac{{ - uk^{1 - \alpha } }}{{(1 - \alpha )(1 + q)}}} \right],
\end{equation}
which is stretched exponential distribution.

For case VI when $\beta=\gamma$, we have
\begin{equation}
\frac{{\partial k_i }}{{\partial t}} = \frac{{pk_i }}{{2t}} + \frac{{2qk_i^\beta  }}{{vt}}.
\end{equation}
According to the derivation in case I, we obtain
\begin{equation}
P(k) \propto \left( {\frac{{vp}}{2}k^{1 - \beta }  + 2q} \right)^{ - \left[ {\frac{2}{{p(1 - \beta )}} + 1} \right]},
\end{equation}
and for large $k$, $P(k) \propto k^{ - \left( {\frac{2}{p} + 1 - \beta } \right)}$.

For case VII when $\alpha=\beta$, we have
\begin{equation}
\frac{{\partial k_i }}{{\partial t}} = \frac{{k_i^\alpha  }}{{ut}} + \frac{{qk_i }}{{2t}}.
\end{equation}
Similarly we obtain
\begin{equation}
P(k) \propto \left( {\frac{{uq}}{2}k^{1 - \alpha }  + 1} \right)^{ - \left[ {\frac{2}{{(1 - \alpha )q}} + 1} \right]},
\end{equation}
and for large $k$, $P(k) \propto k^{ - \left( {\frac{2}{q} + 1 - \alpha } \right)}$.

%For the other cases in Tab. II, qualitatively, case VIII is more homogeneous than case IV for the
%same $p$ and $\beta$, case IX is more homogeneous than case V for
%the same $p$ and $\alpha$, case X is more homogeneous than case VI
%for the same $p$ and $\beta$, case XI is more homogeneous than case
%VII for the same $p$ and $\beta$, and case XII is more homogeneous
%than case VII for the same $p$ and $\alpha$.

The situations in which we can obtain analytical solutions with mean-field method have been shown in Fig. 5.
Bold solid lines mark the situations with power-law degree distribution while bold dashed line indicates the situations with stretched exponential distribution except the two endpoints (exponential for (0, 0, 0) while power law for (1, 1, 1)). Using the common approaches, including mean-field, rate equation and master equation, we cannot obtain all analytical solutions to 27 different scenarios.
\begin{figure}
\centerline{\includegraphics[width=2.5in]{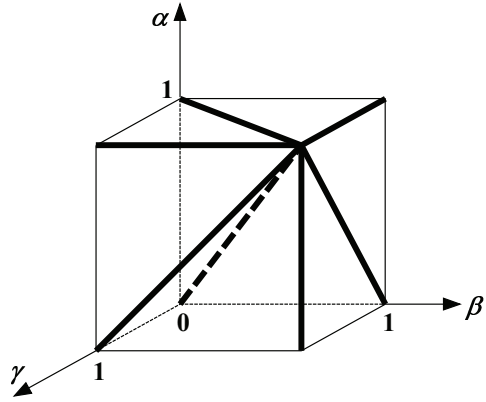}} \caption{The situations which are solvable with mean-field method.}
\label{fig:seven}
\end{figure}

We notice that Eq. (10) can be expressed as
\begin{equation}
\frac{{\partial {k_i}}}{{\partial t}} = \left( {ak_i^\alpha  + bk_i^\beta  + ck_i^\gamma } \right)\frac{1}{t},
\end{equation}
where $a$, $b$ and $c$ are constants. Namely
\begin{equation}
\int\limits_{k_i (t_i )}^{k_i (t)} {\frac{{{\rm{d}}k_i }}{{(ak_i^\alpha   + bk_i^\beta   + ck_i^\gamma  )}}}  = \ln \frac{t}{{t_i }}.
\end{equation}
When $t_j < t_i$,
\begin{equation}
\int_{k_i (t_i )}^{k_i (t)} {\frac{{{\rm{d}}x}}{{ax^\alpha   + bx^\beta   + cx^\gamma  }}}  < \int_{k_j (t_j )}^{k_j (t)} {\frac{{{\rm{d}}x}}{{ax^\alpha   + bx^\beta   + cx^\gamma  }}}.
\end{equation}
Since $k_i (t_i ) = k_j (t_j ) = 1$,
\begin{equation}
\int_1^{k_i (t)} {\frac{{{\rm{d}}x}}{{ax^\alpha   + bx^\beta   + cx^\gamma  }}}  < \int_1^{k_j (t)} {\frac{{{\rm{d}}x}}{{ax^\alpha   + bx^\beta   + cx^\gamma  }}}.
\end{equation}
Thus $k_i (t) < k_j (t)$, i.e. the degrees of the users which appeared in networks before user $i$ are almost everywhere larger than $k_i$. Thus the complementary cumulative degree distribution of networks can be written as
\begin{equation}
P_{\rm{c}} (k) \propto \frac{{N(t_i )}}{{N(t)}} \approx \frac{{t_i }}{t}.
\end{equation}
According to Eqs. (27) and (30), we obtain
%\cite{Guo2010}
\begin{equation}
{P_{\rm{c}}}(k) \propto {{\rm{e}}^{ - \int_1^k {\frac{{{\rm{d}}{k_i}}}{{ak_i^\alpha  + bk_i^\beta  + ck_i^\gamma }}} }}.
\end{equation}

Let $n_1$, $n_2$ and $n_3$ be non-negative integers, $m$ be positive integer, and $\alpha=n_1/m$, $\beta=n_2/m$ and $\gamma=n_3/m$. Further let $s=k_i^{1/m}$ then
\begin{equation}
\int_1^k {\frac{{{\rm{d}}k_i }}{{ak_i^\alpha   + bk_i^\beta   + ck_i^\gamma  }}}  = \int_{1 }^{k^{1/m} } {\frac{{ms^{m - 1} {\rm{d}}s}}{{as^{n_1 }  + bs^{n_2 }  + cs^{n_3 } }}}.
\end{equation}

Suppose that $n_1>n_2>n_3$ and let
\begin{equation}
\frac{{m{s^{m - 1}}}}{{a{s^{{n_1}}} + b{s^{{n_2}}} + c{s^{{n_3}}}}}
=\frac{{m{s^{m - 1-n_3}}}}{{a{s^{{n_1-n_3}}} + b{s^{{n_2-n_3}}} +
c}}= P(s) + \frac{{\hat P(s)}}{{Q(s)}},
\end{equation}
where $P(s)$ and $\hat P(s)$ are polynomials with ${\rm{deg\,}}\hat
P < {\rm{deg\,}}Q$. Furthermore suppose that the polynomial $Q(s)$
has $l$ distinct complex conjugate pairs of roots ${\eta _1} \pm
i{\mu _1}$, $ \ldots $, ${\eta _l} \pm i{\mu _l}$ and $k$ distinct
real roots $\lambda_1$, $ \ldots $, $\lambda_k$, then we have
\begin{equation}
Q(s) = {\prod\limits_{i = 1}^l {\left[ {{{\left( {s - {\eta _i}}
\right)}^2} + \mu _i^2} \right]} ^{{m_i}}}\prod\limits_{i = 1}^k
{{{\left( {s - {\lambda _i}} \right)}^{{n_i}}}},
\end{equation}
where $m_i$ and $n_i$ denote the multiplicities of the roots. For $\hat P(s)/Q(s)$ there
exist real constants $A_{ij}$, $B_{ij}$ and $C_{ij}$ such that
\begin{equation}
\frac{{\hat P(s)}}{{Q(s)}}{\rm{ = }}\sum\limits_{i{\rm{ = }}1}^l
{\sum\limits_{j{\rm{ = }}1}^{{m_i}} {\frac{{{A_{ij}}{\rm{ +
}}{B_{ij}}s}}{{{{\left[ {{{\left( {s{\rm{ - }}{\eta _i}}
\right)}^2}{\rm{ + }}\mu _i^2} \right]}^j}}}} } {\rm{ +
}}\sum\limits_{i{\rm{ = }}1}^k {\sum\limits_{j{\rm{ = }}1}^{{n_i}}
{\frac{{{C_{ij}}}}{{{{\left( {s{\rm{ - }}{\lambda _i}}
\right)}^j}}}} }.
\end{equation}
The second term of the right-hand side of Eq. (35) can easily be
integrated. For the first term when $j=1$ we have
\begin{equation}
\int {\frac{{A{\rm{ + }}Bs}}{{{{\left( {s{\rm{ - }}\eta }
\right)}^2}{\rm{ + }}{\mu ^2}}}} {\rm{d}}s{\rm{ = }}\frac{B}{2}\ln
\left[ {{{\left( {s{\rm{ - }}\eta } \right)}^2}{\rm{ + }}{\mu ^2}}
\right]{\rm{ + }}\frac{{A{\rm{ + }}B\eta }}{\mu }\arctan \left(
{\frac{{s{\rm{ - }}\eta }}{\mu }} \right),
\end{equation}
and when $j>1$
\begin{equation}
\int {\frac{{A + Bs}}{{{{\left[ {{{\left( {s - \eta } \right)}^2} +
{\mu ^2}} \right]}^j}}}} {\rm{d}}s = \frac{{ - B}}{{2(j - 1){{\left[
{{{\left( {s - \eta } \right)}^2} + {\mu ^2}} \right]}^{j - 1}}}} +
\frac{{A + B\eta }}{{{\mu ^{2j - 1}}}}{J_j}\left( {\frac{{s - \eta
}}{\mu }} \right),
\end{equation}
where $J_1(z)=\arctan z$ and
\begin{equation}
{J_{j + 1}}(z) = \frac{z}{{2j{{\left( {{z^2} + 1} \right)}^j}}} +
\frac{{2j - 1}}{{2j}}{J_j}(z).
\end{equation}
Thus according to Eqs. (33)-(38), the primitive function of Eq. (33) can only be the sum of rational functions, logarithmic functions and inverse tangent functions, and for all scenarios in the generalized model, we can analytically obtain their degree distributions though the expressions can be complex in most scenarios.
%we can obtain the analytical
%expressions of degree distributions for the cases in Tab. II though
%the expressions can be complex. In fact the equations also give the
%general solutions to the cases in Tab. I though the analytical expressions
%are simple and have been shown in Tab. I. Therefore

In cases III--XII in Tab. 2, for some special parameters of $\alpha$, $\beta$ or $\gamma$, we can easily obtain the solutions to $P_c(k)$. For example in case VIII, when $\beta=1/3$
\begin{eqnarray}
{P_{\rm{c}}}(k) & \propto & \exp \left[ { - \int_{{1}}^{{k^{1/3}}} {\left( {\frac{{3vs}}{q} - \frac{{3{v^2}}}{{{q^2}p}} + \frac{{{{3{v^3}} \mathord{\left/
 {\vphantom {{3{v^3}} {\left( {{q^2}p} \right)}}} \right.
 \kern-\nulldelimiterspace} {\left( {{q^2}p} \right)}}}}{{qps + v}}} \right){\rm{d}}s} } \right] \nonumber
\\
& = & {\left( {pq{k^{1/3}} + v} \right)^{ - \frac{{3{v^3}}}{{{p^2}{q^3}}}}} \exp \left( { - \frac{{3v{k^{2/3}}}}{{2q}} + \frac{{3{v^2}{k^{1/3}}}}{{{q^2}p}}} \right).
\end{eqnarray}

%in case III, when $\alpha=1/4$, $\beta=1/2$ and $\gamma=3/4$
%\begin{eqnarray}
%{P_{\rm{c}}}(k) & \propto & \exp \left( { - \int_{{1}}^{{k^{1/4}}} {\frac{{4uvw{s^2}{\rm{d}}s}}{{quv{s^2} + quws + pvw}}} } \right) \nonumber
%\\
%& = & \exp \left[ { - \int_{{1}}^{{k^{1/4}}} {\left( {\frac{{4w}}{q} - \frac{{4u{w^2}s + \left( {4v{w^2}p} \right)/q}}{{quv{s^2} + quws + pvw}}} \right){\rm{d}}s} } \right],
%\end{eqnarray}
%where ${\rm{deg\,}}\hat
%P=1$ and ${\rm{deg\,}}Q=2$. According to the following expression ($a \ne 0$)
%\begin{equation}
%\int {\frac{{A + Bs}}{{as^2  + 2bs + c}}} {\rm{d}}s = \left\{
%{\begin{array}{ll}
%   {\frac{B}{{2a}}\ln \left| S \right| + \frac{D}{{a\sqrt \Delta  }}\arctan \left( {\frac{{ax + b}}{{\sqrt \Delta  }}} \right) + C} \hfill & {{\rm{for }}\; \Delta  > 0} \hfill  \\
%   {\frac{B}{{2a}}\ln \left| S \right| + \frac{D}{{2a\sqrt {( - \Delta )} }}\ln \left| {\frac{{ax + b - \sqrt {( - \Delta )} }}{{ax + b + \sqrt {( - \Delta )} }}} \right| + C} \hfill & {{\rm{for }}\; \Delta  < 0} \hfill  \\
%   {\frac{B}{a}\ln \left| ax+b \right| - \frac{D}{{a(ax + b)}} + C} \hfill & {{\rm{for }}\; \Delta  = 0} \hfill  \\
%\end{array}}, \right.
%\end{equation}
%where $S=as^2+2bs+c$, $\Delta=ac-b^2$, $D=aA-bB$ and $C$ is a
%constant, we can get the solution to Eq. (36).

%and for large $k$
%\begin{equation}
%{P_{\rm{c}}}(k)  \propto {k^{ - \frac{{{v^3}}}{{{p^2}{q^3}}}}}{{\mathop{\rm e}\nolimits} ^{ - \frac{{3v{k^{2/3}}}}{{2q}}}}.
%\end{equation}

Although quite controversial online friendship is thought to be
vitally important for the well-being and social capital of people [33, 34]. We use Gini coefficient to quantify the inequality of the degrees of users [35]. Fig. 6 shows the numerical result which is
obtained by averaging over 20 independent realizations. For Eq. (10)
when $\alpha=0.2$, the corresponding numerical result for $0 \le
\beta, \gamma \le 1$ is shown in Fig. 6(a). As expected along minor
diagonal symmetrical pattern emerges. When $\gamma=0.2$ the
corresponding numerical result for $0 \le \alpha, \beta \le 1$ is
shown in Fig. 6(b). The numerical simulations include all cases in
Tab. 2. It is evident that larger preference exponent will result
in greater inequality of the degrees of users and the emergence of
hubs, and thus larger Gini coefficient. Besides we find that from
randomized attachment to PL there is a clear jump for network
heterogeneity, which implies that PL can significantly enhance the
inequality of individual social capital.

\begin{figure}
\centerline{\includegraphics[width=6in]{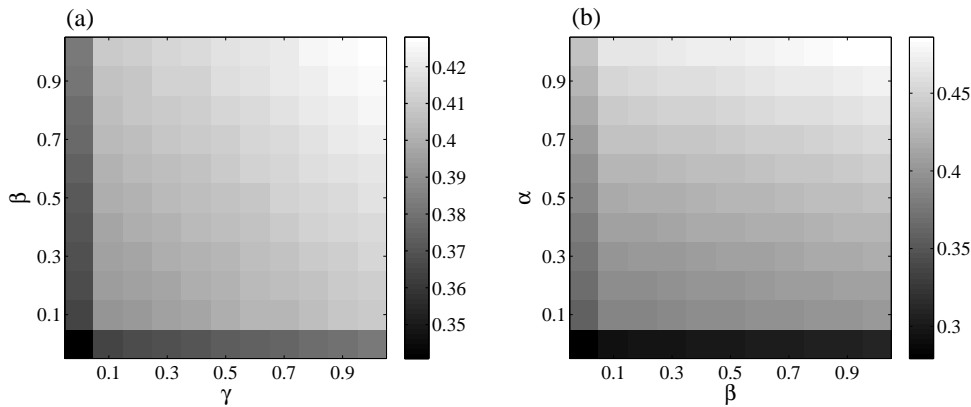}} \caption{Gini coefficients of networks obtained
by numerical simulations with $p = 0.8$ and $N=10^4$.}
\label{fig:seven}
\end{figure}

\section{Conclusion and Discussion}

In summary, we empirically study PL in an evolving OSN, find and validate the existence of linear preference. We propose an analyzable model which reproduces the growth process of the OSN. Furthermore we bring forward a generalized theory of PL and obtain the unified analytical solutions for diverse preference cases with a more general approach.

Why people prefer to attach their links to others who have more
links? Obviously in real life we make friends with someone not
because she/he has many friends but she/he possesses some quality we
expect and is also willing to make friends with us. Thus large
degree predicates that the actor is a worthful and trustworthy
person and making friends with her/him will benefit us. Many
researches have found a positive association between an actor's
degree and that actor's goal achievement, including creativity, job
attainment, professional advancement, political influence and
prestige. Thus a user's degree is a stand-in for her/his true fitness
since direct performance data are costly to gather before the
relationship is made. PL purportedly occurs because actors looking
for new connections use an actor's degree as a proxy for her/his
fitness. A profile owner with many friends will be judged as more
popular than a profile owner with few friends [36].

Kim and Jo proposed several interesting models and explained PA
as rational equilibrium behavior [37]. In fact people are not certain of
the value that they can obtain from forming a link with someone. A
person has an incentive to form a link with another who has many
links because the number of her/his links can convey some
information about her/his value; in an economic sense, the number of
links can be a signal of the value of the person, i.e. the
observable degree contains some information about her/his
unobservable value. From the perspective of economics, if the return
obtained by interacting with someone is greater than the cost, we
like and are willing to continue to maintain this relationship,
especially when the benefit in this relationship outweighs the other
possible relationship. The users with large degrees precisely are
the persons from whom we can expect to get more profit.

PL is widely used as an evolution mechanism of networks. However it
is hard to believe that any individual can get global information
and shape the network architecture based on it. Li et al. found
that the global PA can emerge from the local interaction models,
including the distance-dependent PA evolving model, the acquaintance
network model and the connecting nearest-neighbor model [38]. In fact
Aiello et al. have found that many users join \emph{aNobii} by
creating links to pairs of already connected users [29].

As shown in Fig. 4, the probabilities $p$ and $q$ are time-variant
and cannot be stationary during the real evolution of OSNs. Besides
the activity of users can weaken over time [9]. There
exists a memory kernel which dominates the decline of users'
activity and might be highly skewed, for example obeying power law [39]. Thus a more realistic model can be that $p$,
$q$, $\alpha$, $\beta$ and $\gamma$ in Eq. (10) are all
time-dependent.

Why two people become friends? This question has been widely and
intensively studied in social psychology. Except PL there are
diverse mechanisms which can lead to the formation of dyadic ties,
such as homophily, relational or propinquity mechanisms and physical
attractiveness, and they are intimately interwoven in the evolution
of real social networks and have been found working in the formation
of OSNs [16, 27]. For example homophily has been
found in \emph{Facebook} [40], Microsoft Messenger [41], \emph{LiveJournal} [42], \emph{aNobii} [29], \emph{MySpace} [43] and online dating sites [44, 45]. For relational mechanism, the connecting nearest-neighbor model has been proposed to explain the mechanism [46] and empirical research has shown that this mechanism is at work in \emph{aNobii} [29]. Besides although the Internet transcends some of
the limitations of physical space, proximity still matters in OSNs [47, 48], especially for online dating in which a face-to-face relationship is the goal. PL can account for the degree distribution of
OSNs; however it cannot explain the other structural or sociological
characteristics of the networks. A deeper understanding of these
mechanisms can allow us to better model and predict structure and
dynamics of OSNs [49-51].
Krivitsky et al. made an effort towards the goal [52]. They proposed
a latent cluster random effects model to represent degree
distributions, clustering, and homophily in social networks, however the model is essentially statistical not growing [53].

Most conclusions of the article are theoretical, and need to be
validated by empirical network datasets. Because of the diversity of
purposes of SNSs, there can exist disparate mechanisms dominating
the formation and evolution of OSNs. To the OSNs for general users,
old users can incline to associate with others similar to themselves
and homophily can dominate. While to the OSNs for professionals, old
users can prefer to associate with the celebrities in the same
vocation because personal success in occupation may benefit from the
communication with them. Besides the relative importance of
different mechanisms is also different in different growth stages of
OSNs. In the beginning stage users may incline to establish
friendship relations with the users who are their friends in real
life, while in the later stage users may prefer to make friends with
the users whom they do not know in real life while they are
interested in, which can result in the transition from degree
assortativity to disassortativity [54]. Consider the diversity of users
and the fact that network growth mechanisms tend to be correlated
with each other, for such multidimensional diversity and complexity,
we could only simulate or reproduce one or several of the network
characteristics. Incorporating more social psychological and economic viewpoints
and approaches into the modeling study of OSNs is beneficial to
better understanding the formation of dyadic ties, which will be a
possible future research direction though the analyses would be much
more complex in that setting.

\end{document}